\documentclass[aps,nofootinbib,prd,showpacs,class-pre,a4]{elsarticle} %
\usepackage{amsmath,amssymb,amsfonts,amsthm,bbm}
\usepackage{graphicx}

\usepackage{subeqnarray}

\def\lsim{\hbox{ \raise.35ex\rlap{$<$}\lower.6ex\hbox{$\sim$}\ }}
\def\gsim{\hbox{ \raise.35ex\rlap{$>$}\lower.6ex\hbox{$\sim$}\ }}

\usepackage[T1]{fontenc}					
\usepackage[applemac]{inputenc}				
\usepackage[english]{babel}					
\usepackage{lipsum}						
\usepackage{url} 							
\usepackage{slashed}						
\usepackage{graphicx,setspace}
\usepackage{amsmath, amsfonts, amssymb} 
\usepackage{bbold}

\newcommand{\be}{\begin{equation}}\newcommand{\ee}{\end{equation}}
\newcommand{\bea}{\begin{eqnarray}}\newcommand{\eea}{\end{eqnarray}}
\newcommand{\brr}{\begin{array}}\newcommand{\err}{\end{array}}
\newcommand{\ben}{\begin{enumerate}}\newcommand{\een}{\end{enumerate}}

\newcommand{\ba}{\begin{array}}
\newcommand{\ea}{\end{array}}

\def\lan{\langle}
\def\lf{\left}

\def\non{\nonumber}\def\ran{\rangle}

\def\ri{\right}
\def\al{\alpha}\def\bt{\beta}
\def\de{\delta}\def\ep{\epsilon}
\def\te{\theta}
\def\si{\sigma}
\def\om{\omega}

\def\mass{{_{1,2}}}
\def\flav{{e,\mu}}\def\1{{_{1}}}\def\2{{_{2}}}
\def\bk{{\bf {k}}}

\newcommand{\ide}{1\hspace{-1mm}{\rm I}}

\usepackage{xcolor}

\begin{document}

\title{On the r$\mathrm{\hat{o}}$le of rotations and Bogoliubov transformations in neutrino mixing}

\author[UniSA,INFN]{M.~Blasone}
    \ead{blasone@sa.infn.it}
\author[UniSA,INFN]{M.V.~Gargiulo}
    \ead{mgargiulo@unisa.it}
\author[UniSA,INFN]{G.~Vitiello}
    \ead{vitiello@sa.infn.it}
\address[UniSA]{Dipartimento di Fisica, Universit\`a di Salerno, Via Giovanni Paolo II, 132 84084 Fisciano, Italy}
\address[INFN]{INFN Sezione di Napoli, Gruppo collegato di Salerno, Italy}

\begin{abstract}
We show that mixing transformations for Dirac fields arise as a consequence of the non-trivial interplay between rotations and Bogoliubov transformations at level of ladder operators. Indeed the non-commutativity between the algebraic generators of such transformations turns out to be responsible of the  unitary inequivalence of the flavor and mass representations and of the associated vacuum structure. A possible thermodynamic interpretation is also investigated.
\end{abstract}


\maketitle

\section{Introduction}
Since Pontecorvo's pioneering work \cite{Pontecorvo} the theoretical basis of neutrino mixing has been studied in great detail \cite{Giuntibook} and a quantum field theory (QFT) formalism for mixed fields has been developed \cite{BV95,Massimo2,Fujii,Ji,Massimo3,Mavromatos,Massimo5,Massimo6,rindmix,bjvbook}. Phenomenological and experimental developments have successfully confirmed \cite{Super_Kamiokande_Collaboration,SNO_Collaboration,KamLAND_Collaboration,GNO_GALLEX,atmospheric,OPERA,RevNOsc} the original proposal of the occurrence of the phenomenon of neutrino mixing and oscillations, thus opening new scenarios beyond the Standard Model (SM) of elementary particle physics. Puzzling questions remain, however, open. Among these, the problem of the origin of the non vanishing neutrino masses and mixings is a crucial one.

The QFT formalism has shown the limits of the quantum mechanical approximation in the treatment of mixing of neutrino fields by exhibiting the unitary inequivalence of the vacuum for neutrino fields with definite flavor (flavor vacuum) and the ones with definite mass.  The unitary inequivalence between representations of the canonical (anti-)commutation relations is a characteristic feature of QFT, which is absent in quantum mechanics (QM) due to the von Neumann theorem \cite{vonNeumannQMbook}. It has been shown \cite{QFTformalism} that many physically relevant aspects in the mixing and oscillation phenomenon are indeed consequences of such a QFT characteristic feature.

In this paper we focus on the algebraic structure of the field mixing generator. In QM the mixing transformation looks like a rotation operating on massive neutrino states. We show explicitly that such a rotation is not sufficient for implementing the mixing transformation at level of fields. It is necessary, in fact, also the action of a Bogoliubov transformation which operates a suitable mass shift. Such a property of Bogoliubov transformations has been already known and used since long time \cite{Ume1,Ume,Ume2,UTK}, e.g. in renormalization theory or in the dynamical generation of mass \cite{UTK,Miransky}. Bogoliubov transformations are also used in recent studies of neutrino mixing in astrophysics \cite{Volpe}. The key point in our analysis is the non commutativity between rotation and Bogoliubov transformations, a feature which turns out to be at the origin of the inequivalence among mass and flavor vacua.

The paper is organized as follows. In Section II we investigate the compatibility of the mixing transformation at level of states and fields, and show that a Bogoliubov transformations is required. In Section III we analyze the condensate nature of the flavor vacuum and the r$\mathrm{\hat{o}}$le played by the non commutativity between the rotation and the Bogoliubov transformation. The possibility of a thermodynamical interpretation of such a condensate is considered in Section IV. Finally, in Section V we draw our conclusions. The paper is completed with three Appendices.

\section{Rotation and Bogoliubov transformations}

Pontecorvo mixing transformations are written as a rotation of the states with definite masses $|\nu_1\rangle$, $|\nu_2\rangle$, into those with definite flavor $|\nu_e\rangle$ and $|\nu_\mu\rangle$ as \cite{Pontecorvo}
\bea\label{Pontmix1}
|\nu_e\rangle&=&\cos{\theta} \,|\nu_1\rangle + \sin{\theta} \,|\nu_2\rangle,
\\ \label{Pontmix2}
|\nu_\mu\rangle&=&\cos{\theta}\, |\nu_2\rangle - \sin{\theta} \, |\nu_1\rangle.
\eea
On the other hand, Standard Model is formulated in terms of fields\footnote{Our analysis is limited to the case of two Dirac neutrinos. Extension to three neutrinos is in our plans. However, we have good reasons to believe that the present results are general, since our arguments are of algebraic nature.} and there neutrino mixing appears in the following form \cite{ChengLi}
\bea
\label{fieldmix1}
\nu_{e}(x) &=&  \cos\te \;\nu_{1}(x)  \, +\,\sin\te\; \nu_{2}(x) \, ,
\\ \label{fieldmix2}
\nu_{\mu}(x) &=& \cos\te\;\nu_{2}(x)\, - \, \sin\te \;\nu_{1}(x)  \, ,
\eea
where $x\equiv ({\bf x}, t)$. The generator of such a transformation is \cite{BV95}
 \be
 \label{generatorG} G(t; \theta, m_1,m_2) = \exp\lf\{\te \int d^{3}{\bf x} \lf(\nu_{1}^{\dag}(x) \nu_{2}(x) - \nu_{2}^{\dag}(x) \nu_{1}(x) \ri)\ri\}\;.
\ee
 The question then arise to what extent the two above transformations are equivalent. It has been shown \cite{BV95} that this is not the case and indeed a deep conceptual difference is present between mixing of states and mixing of fields.
The results also extend to the mixing phenomenon of any particle, and are not limited to the case of Dirac neutrinos.

Let us now consider the expansion for the Dirac fields  $\nu_{1}$ and $\nu_{2}$ with definite masses appearing in Eqs.(\ref{fieldmix1}),(\ref{fieldmix2}):
\bea
\label{eqn:two}
\nu_{i}(x) = \sum_{r} \int \frac{d^3 \bk}{(2\pi)^\frac{3}{2}}   \lf[u^{r}_{{\bf k},i}(t) \al^{r}_{{\bf k},i}\:+ v^{r}_{-{\bf k},i}(t) \bt^{r\dag }_{-{\bf k},i} \ri] e^{i {\bf k}\cdot {\bf x}},\quad i=1,2,
\eea
where $u^{r}_{{\bf k},i}(t)=e^{-i\om_{k,i} t} u^{r}_{{\bf k},i}$ and $v^{r}_{-{\bf k},i}(t)=e^{i\om_{k,i} t}v^{r}_{-{\bf k},i}$, with $\om_{k,i}=\sqrt{{\bf k}^2+m_i^2}$. The $\al^{r}_{{\bf k},i}$ and the $\bt^{r }_{-{\bf k},i}$ ($r=1,2$), are the annihilation operators for the vacuum state $|0\ran_{1,2}\equiv|0\ran_{1}\otimes |0\ran_{2}$. See Appendix for other useful relations.

Observe that Eqs.(\ref{Pontmix1}),(\ref{Pontmix2}) can be seen as arising by the application to the vacuum state $|0\ran_{1,2}$ of the rotated operators:
\bea
\label{rota1}\hspace{-8mm}
R(\theta)^{-1}\alpha^{r\dagger}_{{\bf k},1} R(\theta)
&=& \cos\te\, \alpha^{r\dagger}_{{\bf k},1} \,+\,  e^{-i\psi_k}\, \sin\te \, \alpha^{r\dagger}_{{\bf k},2},
\\ [2mm]\label{rota2}\hspace{-8mm}
R(\theta)^{-1}\alpha^{r\dagger}_{{\bf k},2} R(\theta)
&=& \cos\te\, \alpha^{r\dagger}_{{\bf k},2} \,-\,  e^{i\psi_k}\, \sin\te \, \alpha^{r\dagger}_{{\bf k},1},
\eea
and similar ones for $\beta^{r\dagger}_{{\bf k},i}$. An arbitrary phase $\psi_k$ has been also included.
The generator $R(\te)$  is indeed the one of a  rotation:
\bea
\label{rotgen} \non\hspace{-9mm}
R(\theta)&=&\exp\lf\{    \theta  \sum_r \int \frac{d^3 \bk}{(2\pi)^\frac{3}{2}} \Big[\Bigl(\alpha^{r\dagger}_{{\bf k},1}\alpha^{r}_{{\bf k},2}+
\beta^{r\dagger}_{- {\bf k},1} \beta^{r}_{-{\bf k},2} \Bigr) e^{i\psi_k}\ri.\\  \hspace{-9mm}
&-&\lf. \Bigl(\alpha^{r\dagger}_{{\bf k},2}\alpha^{r}_{{\bf k},1} +
\beta^{r\dagger}_{-{\bf k},2} \beta^{r}_{-{\bf k},1} \Bigr) e^{-i\psi_k} \Big]\ri\} ,
\eea
Notice that the unitary operator $R^{-1}=R^\dag$ leaves the vacuum invariant:
\bea
\label{rotvac}
R^{-1}(\theta)  |0\ran_{1,2} = |0\ran_{1,2} \; .
\eea

In order to study the generator
$G(t;\theta, m_1, m_2)$, Eq.(\ref{generatorG}), it is useful to introduce another canonical transformation, the Bogoliubov transformation:
\bea
\label{tialpha} \hspace{-8mm}
\tilde{\alpha}^{r\dagger}_{{\bf k},i} &\equiv & B^{-1}_i(\Theta_i)  \, \alpha^{r\dagger}_{{\bf k},i} \,  B_i(\Theta_i)  = \cos{\Theta_{{\bf k},i}}\,\alpha^{r\dagger}_{{\bf k},i} - \epsilon^r\,e^{i\phi_{k,i}}\, \sin{\Theta_{{\bf k},i}}
\,\beta^{r}_{-{\bf k},i},\\[2mm] \hspace{-8mm}
\label{tibeta} \hspace{-8mm}
\tilde{\beta}^{r\dagger}_{-{\bf k},i}&\equiv & B^{-1}_i(\Theta_i) \,
\beta^{r\dagger}_{-{\bf k},i}  \, B_i(\Theta_i)  = \cos{\Theta_{{\bf k},i}}\,\beta^{r\dagger}_{-{\bf k},i} + \epsilon^r \,e^{-i\phi_{k,i}}\,\sin{\Theta_{{\bf k},i}}
\, \alpha^{r}_{{\bf k},i},
\eea
with $i=1,2$ and the generator(s)
\bea
\label{boggen}  \hspace{-5mm}
B_i(\Theta_i)=\exp{\Biggl\{  }   \sum_r \int \frac{d^3 \bk}{(2\pi)^\frac{3}{2}} \Theta_{{ \bf k},i}
\;\epsilon^r  \Big[  \alpha^{r}_{{\bf k},i}\beta^{r}_{-{\bf k},i} e^{-i\phi_{k,i}} - \beta^{r\dagger}_{-{\bf k},i} \alpha^{r\dagger}_{{\bf k},i}e^{i\phi_{k,i} }\Big]\Biggr\} .
\eea
Since $[ B_1(\Theta_1) ,  B_2(\Theta_2) ]=0$, we may also define
$
B(\Theta_1,\Theta_2) \equiv B_2(\Theta_2) B_1(\Theta_1).
$
Note that,   the Bogoliubov transformation does not leave invariant the vacuum~$|0\ran_{1,2}$. Defining $| \widetilde{0}\ran_{1,2} \equiv  B^{-1}(\Theta_{1},\Theta_{2})|0\ran_{1,2}$, we have
\bea
\label{eqn:BminusO}  
| \widetilde{0}\ran_{1,2}=  \prod_{i=1,2} \prod_{{\bf k},r} \Big[ \cos{\Theta_{{\bf k},i}} +
\epsilon^r e^{i\phi_{k,i} }\sin{\Theta_{{\bf k},i}} \alpha^{r \dagger}_{{\bf k},i} \beta^{r \dagger}_{-{\bf k},i} \Big] |0\ran_{1,2}.
\eea
The states $| \widetilde{0}\ran_{1,2} $ and $| 0\ran_{1,2} $ become orthogonal in the infinite volume limit,  thus giving rise to inequivalent representations \cite{bjvbook}. This is a well-known feature of QFT \cite{Ume2} reflecting into the non-unitary nature (in the infinite volume limit) of the generator of Bogoliubov transformations.

We now consider the action of the  rotation  Eq.(\ref{rotgen})  on the fields $\nu_1$ and $\nu_2$:
\bea
\non
&&\hspace{-10mm}R^{-1}(\te)\nu_1(x)R(\te)=\cos\te\,\nu_1(x)\\
&& \hspace{-10mm}\qquad\qquad+\sin\te \sum_r \int \frac{d^3 \bk}{(2\pi)^\frac{3}{2}}\,e^{i {\bf k}\cdot{\bf x}} \lf(e^{i \psi_k}\alpha_{{\bf k},2}^r \,u_{{\bf k},1}^r (t)+ e^{-i \psi_k}\beta^{r\dagger}_{{\bf k},2} \,v_{-{\bf k},1}^r (t) \ri) ,
\\ \non
&&\hspace{-10mm}R^{-1}(\te)\nu_2(x)R(\te)=\cos\te\,\nu_2(x)\\
&& \hspace{-10mm}\qquad\qquad-\sin\te \sum_r \int \frac{d^3 \bk}{(2\pi)^\frac{3}{2}}\,e^{i {\bf k}\cdot{\bf x}}  \lf(e^{- i \psi_k}\alpha_{{\bf k},1}^r \,u_{{\bf k},2}^r (t)+ e^{i \psi_k}\beta^{r\dagger}_{{\bf k},1} \,v_{-{\bf k},2}^r (t) \ri) .
\eea
The above expressions do not fully reproduce the mixing at level of fields, cf.Eqs.(\ref{fieldmix1}),(\ref{fieldmix2}): the problem is that the last term in  the r.h.s. of these equations appears as the expansion of the field in the ``wrong'' basis. However,  it is possible to recover the wanted expression by means of a suitable Bogoliubov transformation, which implements a mass shift. Let us see this for the field $\nu_1$:
\bea
 \non \hspace{-8mm}
&{}&B_2^{-1}(\Theta_2)\,R^{-1}(\te)\, \nu_1(x)\,R(\te)\,B_2(\Theta_2)=\\ \non
 \hspace{-8mm}
&{}&= \cos\te\,\nu_1(x)+\sin\te \sum_r \int \frac{d^3 \bk}{(2\pi)^\frac{3}{2}}e^{i {\bf k}\cdot{\bf x}}\lf( e^{i \psi_k}\tilde{\alpha}_{{\bf k},2}^r u_{{\bf k},1}^r(t) + e^{- i \psi_k} \tilde{\beta}^{r \dagger}_{{\bf k},2} v_{-{\bf k},1}^r (t) \ri)\qquad\\
 \hspace{-8mm}
&{}&=\cos\te\,\nu_1(x)+\sin\te \sum_r \int \frac{d^3 \bk}{(2\pi)^\frac{3}{2}}e^{i {\bf k}\cdot{\bf x}} \lf(e^{i \psi_k}\alpha_{{\bf k},2}^r \hat{u}_{{\bf k},1}^r (t)+ e^{- i \psi_k}\beta^{r \dagger}_{{\bf k},2} \hat{v}_{{-\bf k},1}^r (t) \ri)\, ,\qquad
\eea
where
\bea \hspace{-11mm}
&{}&\hat{u}_{{\bf k},1}^r (t)  =  u_{{\bf k},1}^r e^{-i\om_{k,1} t}e^{i \psi_k}  \cos\Theta_{{\bf k},2} +\epsilon^r v_{-{\bf k},1}^r e^{i\om_{k,1} t } e^{- i \phi_{k,2} } e^{-i \psi_k} \sin\Theta_{{\bf k},2}\, ,\\[2mm]
 \hspace{-11mm}
&{}&\hat{v}_{-{\bf k},1}^r (t) =  v_{-{\bf k},1}^r e^{i\om_{k,1} t} e^{-i \psi_k} \cos\Theta_{{\bf k},2}-\epsilon^r u_{{\bf k},1}^r e^{-i\om_{k,1} t} e^{-i\phi_{k,2}} e^{-i \psi_k} \sin\Theta_{{\bf k},2}\, .
\eea
For $\hat{\Theta}_{{\bf k},2}= \cos^{-1}\lf(e^{-i \psi_k}U_{{\bf k}}(t)\ri)$, with $U_{{\bf k}}(t)\equiv u^{r\dag}_{{\bf k},2}(t)u^{r}_{{\bf k},1}(t)$ (see Appendix), the Bogoliubov transformation $B_2(\hat{\Theta}_2)$ produces  the mass shift  $m_2-m_1$, such that\footnote{An equivalent choice is $\hat{\Theta}_{{\bf k},2}=\sin^{-1}\lf( e^{i\phi_{k,2}} e^{i \psi_k} V_{{\bf k}}(t)\ri)$ with $V_{{\bf k}}(t)\equiv \ep^{r}\;u^{r\dag}_{{\bf k},1}(t)v^{r}_{-{\bf k},2}(t)$.}
$
\hat{u}_{{\bf k},1}^r (t) =  u_{{\bf k},2}^r (t)\, $ and $
\hat{v}_{-{\bf k},1}^r (t)=  v_{-{\bf k},2}^r (t).
$
In definitive, the action of $B_2^{-1}(\hat{\Theta}_2)\,R^{-1}(\te)$ produces the desired transformation of the field $\nu_1$, cf. Eq.(\ref{fieldmix1}). A similar reasoning can be done for $\nu_2$, using $B_1^{-1}(\hat{\Theta}_1)\,R^{-1}(\te)$, with $\hat{\Theta}_{{\bf k},1}= \cos^{-1}\lf(e^{i \psi_k}U_{{\bf k}}(t)\ri)$.

Note that the r$\mathrm{\hat{o}}$le of the Bogoliubov transformation in the process of (dynamical) mass generation is well known, see for example Refs.\cite{UTK,Miransky}.

\section{Vacuum structure and non-commutativity}

In the previous Section, we have shown the incompatibility of the mixing transformation as mere rotations both for states and fields, and the necessity of implementing a mass shift for reproducing the correct relations for fields: such an operation is highly non-trivial and indeed requires infinite energy (in the infinite volume limit).

On the other hand, the results of Section II are incomplete in that two different generators are needed for $\nu_1$ and $\nu_2$, whereas we know the algebraic generator for fields to be that of Eq.(\ref{generatorG}). It thus arises the problem of the decomposition  of such generator in terms of rotation and Bogoliubov transformations; a preliminary solution to this problem has been presented in \cite{ProcDice15}.
The full decomposition of the mixing generator is given by (see Appendix)
\bea\label{gendec1}
G(t;\theta, m_1, m_2)= B^{-1}(t;m_1,m_2)\,  R(t;\te) \, B(t;m_1,m_2)\, ,
\eea
where the notation is now $f(\Theta_i(m_i))\equiv f(m_i)$; $R(t;\te)$ and $B(t;m_1,m_2)$ are defined as in Eqs.(\ref{rotgen}),(\ref{boggen}), with the phases $\phi_{k,i} \equiv 2 \omega_{k,i} t $ and $ \psi_k \equiv  (\omega_{k,1} - \omega_{k,2}) t $ and the condition  $\Theta_{{\bf k},i} = \frac {1} {2} \cot^{-1} (\frac{|{\bf k}|}{m_i})$ has been used \cite{ProcDice15} .
From Eq.(\ref{gendec1}) it appears evident that the difference between $G$ and $R$  relies in the non zero value of the commutator $[R,B]$.

The explicit form of $G(\theta)$ in terms of ladder operators  is given by Eq.(\ref{genexp}) in the Appendix.
It is possible to rewrite $G(\theta)$  (at $t=0$) as
\bea\label{GeneratorJ}
&&\hspace{-8mm}G(\te)=\exp\Biggl\{ 2 \theta \sum_{r}\int\frac{d^3 \bk}{(2\pi)^\frac{3}{2}} \Big[U_{\bf k} J^r_{{\bf k},3} - \epsilon^{r}V_{\bf k} J^r_{{\bf k},2} \Big]\Biggr\}\, ,
\eea
where we have introduced the following operators\footnote{We also have $J^r_{{\bf k},1} \equiv \frac 1 2( K^r_{{\bf k},1} - K^r_{{\bf k},2}) $ with $ K^r_{{\bf k},i} \equiv   \alpha^{r}_{{\bf k},i}\beta^{r}_{-{\bf k},i} -\beta^{r \dagger}_{-{\bf k},i} \alpha^{r \dagger}_{{\bf k},i}$ and $\ln B_i(\Theta_{{\bf k},i})= \int \frac{d^3 \bk}{(2\pi)^{3/2}}\Theta_{{\bf k},i} \sum_r K^r_{{\bf k},i}$ ; $\ln R(\theta)= \,2 \theta \int \frac{d^3 \bk}{(2\pi)^{3/2}}\, \sum_r J^r_{{\bf k},3}$.}:
\bea
\label{Jeqn}  \hspace{-8mm}
J^r_{{\bf k},1} &&\equiv  \frac 1 2 \Big[ (\alpha^{r}_{{\bf k},1}\beta^{r}_{-{\bf k},1} -\beta^{r \dagger}_{-{\bf k},1} \alpha^{r \dagger}_{{\bf k},1} ) -( \alpha^{r}_{{\bf k},2}\beta^{r}_{-{\bf k},2} -\beta^{r \dagger}_{-{\bf k},2} \alpha^{r \dagger}_{{\bf k},2}) \Big], \\
 \hspace{-8mm} 
J^r_{{\bf k},2}  &&\equiv- \frac  1 2 \Big[ (\alpha^{r}_{{\bf k},1}\beta^{r}_{-{\bf k},2} -\beta^{r \dagger}_{-{\bf k},2} \alpha^{r \dagger}_{{\bf k},1} )+( \alpha^{r}_{{\bf k},2}\beta^{r}_{-{\bf k},1} -\beta^{r \dagger}_{-{\bf k},1} \alpha^{r \dagger}_{{\bf k},2}) \Big], \\
\hspace{-8mm}
J^r_{{\bf k},3}  &&\equiv  \frac 1 2 \Big[ (\alpha^{r \dagger}_{{\bf k},1}\alpha^{r}_{{\bf k},2} + \beta^{r \dagger}_{-{\bf k},1} \beta^{r}_{-{\bf k},2}) - (\alpha^{r \dagger}_{{\bf k},2}\alpha^{r}_{{\bf k},1} + \beta^{r \dagger}_{-{\bf k},2} \beta^{r}_{-{\bf k},1})  \Big] ,
\eea
which close the $su(2)$ algebra: $[J^r_{{\bf k},i},J^r_{{\bf k},j}]=\epsilon_{ijk} J^r_{{\bf k},k}$ with $i,j,k=1,2,3$.
Moreover, considering that the Bogoliubov coefficients $U_{\bf k}$ and $V_{\bf k}$ appearing in Eq.(\ref{GeneratorJ}) can be written as $ U_{\bf k} \, = \, \cos(\Theta_{{\bf k},2} - \Theta_{{\bf k},1})\, $, $ V_{\bf k} \, = \, \sin(\Theta_{{\bf k},2} - \Theta_{{\bf k},1})$, in the limit of small $(\Theta_{{\bf k},2} - \Theta_{{\bf k},1})$, it is possible to expand $V_{\bf k}$ in terms of the adimensional parameter $a\equiv \frac{(m_2 - m_1)^2}{m_1 m_2}$ so that $ U_{\bf k} \, \cong 1 \; $, $ V_{\bf k} \, \cong  a \tilde{V}_{\bf k} $,
up to $o[(a)^2]$ where $\tilde{V}_{\bf k} \equiv \frac {|\bk| \sqrt{m_1 m_2}}{2(|\bk|^2 + m_1 m_2 )}$ and thus,
\bea
\label{mixgenapproxI}
 G(\te) &\cong &  1 \hspace{-1 mm} \text{I} + 2 \theta \int \frac{d^3 \bk}{(2\pi)^\frac{3}{2}}\, \sum_r J^r_{{\bf k},3} + 2 \theta \, a \, \int \frac{d^3 \bk}{(2\pi)^\frac{3}{2}}  \,\tilde{V}_{\bf k}\,   \sum_r \epsilon^r J^r_{{\bf k},2}\, .
\eea
It is easy to see as this generator becomes the identity when $\theta=0$ and is equivalent to a mere rotation when $a=0$, i.e. $m_2 = m_1$. Moreover, the last term shows the explicit dependance on the true physical parameters of the mixing transformation, i.e. $\theta$ and $a$. Notice that the adimensional parameter $a$ appears at second order in the expansion, being linked with the commutator $J^r_{{\bf k},2}=[J^r_{{\bf k},3},J^r_{{\bf k},1}]$ which can be interpreted as a \emph{non-diagonal Bogoliubov transformation}, and is the first non trivial term which contributes to the flavor vacuum structure\footnote{ The complete operatorial structure of the flavor vacuum (Eq.(\ref{flavorvacuum}) in the Appendix) is obtained already at the second order approximation.}.
This feature can be further understood by looking at the tilde vacuum, defined as (cf. Eq.(\ref{eqn:BminusO})):
\bea 
  |\tilde{0}\ran_{1,2}\cong \Biggl[ 1 \hspace{-1 mm} \text{I}  +   \, \int \frac{d^3 \bk}{(2\pi)^\frac{3}{2}} \,\sum_r \Big(\Theta_{{\bf k},1} \, \alpha^{r \dagger}_{{\bf k},1} \beta^{r \dagger}_{-{\bf k},1} + \Theta_{{\bf k},2}  \,\alpha^{r \dagger}_{{\bf k},2} \beta^{r \dagger}_{-{\bf k},2} \Big)\Biggr]|0\ran_{1,2}\, ,
\eea
for $\Theta_{{\bf k},i} $ small, and comparing it with the flavor vacuum $|0\ran_\flav \equiv G^{-1}(\te)|0\ran_{1,2}$ obtained in our approximation:
\bea
\label{vacuumaprox1}
|0\ran_\flav \cong \Biggl[ 1 \hspace{-1 mm} \text{I}  + \, \theta \, a \, \int \frac{d^3 \bk}{(2\pi)^\frac{3}{2}} \, \,\tilde{V}_{\bf k}\, \sum_r \epsilon^r  \Big( \alpha^{r \dagger}_{{\bf k},1}\beta^{r \dagger}_{-{\bf k},2}+ \alpha^{r \dagger}_{{\bf k},2} \beta^{r \dagger}_{-{\bf k},1}\Big) \Biggr]|0\ran_{1,2}\, .
\eea
Notice that, although the operatorial structure of the two above equations is similar, Eq.(\ref{vacuumaprox1}) exhibits non diagonal operatorial terms.
From Eq.(\ref{vacuumaprox1}) we see that $|0\ran_\flav$ cannot be reduced as a tensor product of vectors built on  $|0\ran_{1,2}$: this indeed confirms that the phenomenon of flavor mixing is related to the entanglement of mass eigenstates (see \cite{entangoscill} for the discussion of entanglement in the context of particle mixing and oscillations).
Another interesting feature of this phenomenon appears as one analyses more closely the parameter $a$, which in order to exist needs at least two fermion families to be present. In fact, with just one family the only adimensional parameter one can form is $\frac{|\bk|}{m}$, which however depends on $k$ and thus cannot be extracted from the integrals.

Finally, let us express the flavor vacuum by means of the full finite decomposition in Eq.(\ref{gendec1}):
\bea
\label{eqn:twotwo}  \hspace{-5mm}   |0\ran_\flav \,= \, |0\ran_{1,2}\; + \; \Big[B(m_1,m_2)  \; ,
\; R^{-1}(\theta)\Big] \;  |\widetilde{0}\ran_{1,2}\, ,
\eea
where $ |\widetilde{0}\ran_{1,2}$ is defined in Eq.(\ref{eqn:BminusO}). We, thus, see how a condensate nature, made of particle-antiparticle pairs with same or different masses \cite{BV95}, arises as a consequence of the non vanishing commutator $[B, R^{-1}]$.
Indeed, a condensate is already present in the Bogoliubov vacuum $| \widetilde{0}\ran_{1,2} $, for which it is possible to compute a condensation density:
\bea\label{vuotoBOG}  \hspace{-5mm}  \,_{1,2}\lan \widetilde{0}| \al_{{\bf k},i}^{r \dag}
\al^r_{{\bf k},i} | \widetilde{0}\ran_{1,2} =\;_{1,2}\lan\widetilde{0}| \bt_{-{\bf k},i}^{r \dag}
\bt^r_{-{\bf  k},i} | \widetilde{0}\ran_{1,2} =  \sin^2\Theta_{{\bf k},i},
\eea
with $ i=1,2$.
The condensation density of the flavor vacuum differs from the one of the Bogoliubov vacuum and is given by
\bea\label{seventeen}  \hspace{-10mm} \non
\;_{e,\mu}\lan 0(t)| \al_{{\bf k},i}^{r \dag} \al^r_{{\bf k},i} |0(t)\ran_{e,\mu}&=&\;_{e,\mu}\lan 0(t)| \bt_{-{\bf k},i}^{r \dag} \bt^r_{-{\bf  k},i} |0(t)\ran_{e,\mu}\\
 \hspace{-10mm} &=&\sin^{2}\te\; \sin^2({\Theta_{{\bf k},1} - \Theta_{{\bf k},2}}),
\eea
with $ i=1,2$.
We stress that, such condensation density, vanishes when either $\te=0$ and/or $m_1=m_2$, which are the cases in which there is no mixing.

Withal, as a result of the non vanishing commutator in Eq.(\ref{eqn:twotwo}) , one finds a gap in the vev of the 
energy on the two vacua $ \Delta E_\bk \equiv \,_{e,\mu}\lan0| H_{\bf k} |0 \ran_{e, \mu} - \,_{1,2}\lan 0| H_{\bf k} |0\ran_{1,2} $ :
\bea \label{engap}
\Delta E_\bk = 2 (\omega_{ k,1} + \omega_{k,2})  \sin^2{\theta} \sin^2({\Theta_{{\bf k},1} - \Theta_{{\bf k},2}})\, ,
\eea
where $H_{\bf k}\equiv H_{{\bf k},1} + H_{{\bf k},2} $. A detailed analysis of the energy gaps among the vacua $|0 \ran_{e, \mu}$, $| \widetilde{0}\ran_{1,2}$ and  $|0\ran_{1,2} $ is given in \cite{ProcDice15}.

\section{Thermodynamical properties}
\label{sect4}

In this Section we investigate the possibility of a thermodynamical interpretation for the condensate structure of the flavor vacuum. We proceed in analogy with Thermo Field Dynamics (TFD) for fermions, where a thermal vacuum is generated by means of a suitable Bogoliubov transformation:
\bea
\label{TFDvac}
|0(\vartheta)\rangle  =  \prod_{{\bf k},r} \Big[ \cos{\vartheta_{{\bf k}}} + \,\sin{\vartheta_{{\bf k}}}\, \alpha^{r \dagger}_{{\bf k}} \tilde{\alpha}^{r \dagger}_{{\bf k}} \Big] |0\ran_{1,2},
\eea
where $\alpha$ and ${\tilde \al}$ are fermion operators, and $\vartheta = \vartheta(\beta)$. Note that a  ``fictitious'' system (the \emph{tilde} system), with the same structure of the physical system, is introduced and is interpreted as a thermal bath. 
According to \cite{Ume1}, such a state can be written as
 \bea
 |0(\vartheta)\rangle = \exp\left(-\frac{S_{\al}}{2}\right)|I\rangle=\exp\left(-\frac{S_{\tilde{\al}}}{2}\right)|I\rangle
 \eea
 with $|I\rangle \equiv \exp\left(\sum_{\bk,r}\tilde{\al}_{{-\bf k}}^{r\dag}\al_{{\bf k}}^{r\dag}\right)|0\rangle$, and
  \bea S_{\al}= -\sum_{\bk, r} \left(\al_{{\bf k}}^{r\dag}\al_{{\bf k}}^{r}
\ln\sin^{2}\vartheta_{{\bf k}} + \al_{{\bf k}}^{r}\al_{{\bf k}}^{r\dag}\ln\cos^{2}\vartheta_{{\bf k}}\right).
\eea
In the above derivation one makes use of the following relations
\bea
\label{relentrTFD}  \hspace{-8mm}
e^{-\frac{S_{a}}{2}} \al_{{\bf k}}^{\dag} e^{\frac{S_{\al} }{2}} = \tan{\vartheta_{{\bf k}}} \al_{{\bf k}}^{\dag}\,, \quad\quad\quad
e^{-\frac{S_{a}}{2}} \tilde{\al}_{{\bf k}}^{\dag} e^{\frac{S_{\al} }{2}} =  \tilde{\al}_{{\bf k}}^{\dag}\, .
\eea
A similar expression holds for $S_{\tilde{\al}}$. $S_{\al}$ (or $S_{\tilde{\al}}$) can, thus, be interpreted as the entropy function associated to the vacuum condensate.
We also have\footnote{we use the notation $ \langle  0(\vartheta)| * |0(\vartheta)\rangle  \equiv  \langle * \rangle_{\vartheta}$.}
\bea \label{TFDcond}
n_k \equiv \langle \alpha_{{\bf k}}^{r\dag}\alpha_{{\bf k}}^{r} \rangle_{\vartheta} =  \sin^2({\vartheta_{\bf k}})\, .
\eea
The expectation value of the Hamiltonian $H_\al=\sum_{\bf k}{ \epsilon_k \alpha_\bk^{\dag} \alpha_\bk }$ is $\langle H_\al \rangle_{\vartheta}  =  \sum_k{ \epsilon_k n_k}$. We will use $\omega_k = \epsilon_k - \mu$, with $\mu$ being the chemical potential.
The vev on the thermal vacuum of the entropy is: $\langle S_{\al} \rangle_{\vartheta} = -2\sum_{{k}}\big( n_{k}\ln n_{k}+ (1-n_{k}) \ln(1-n_{k})\big)$.
One also considers the following quantity: $\Omega= \langle H_\al  - \frac 1 \beta S_{\al} - \mu N_\al \rangle_{\vartheta}$ ,
which can be identified as a thermodynamical potential \cite{Ume1}. Extremization of $\Omega $ with respect to $\vartheta_\bk$  leads to the Fermi-Dirac distribution.
\bea
\label{thetabogferm}
n_k =\frac {1}{e^{\beta\omega_{ k}}+1}.
\eea

We apply a similar reasoning of the one in \cite{Ume1}, also for the case of the flavor vacuum generated by $G_{t}(\theta, m_1,m_2) $ as in Eq.(\ref{generatorG}) and assume that it is possible to rewrite it as:
\bea
\label{entrinv}
|0\rangle_{e, \mu} = e^{- \frac{S^f_{i}} {2}} |I_f\rangle \, ,
\eea
where  $i=1,2$, $f$ denotes ``flavor'', and\footnote{Here we have chosen to separate the physical and tilde systems (in analogy with TFD) according to the mass index.
}
$S^f_{i}\equiv \sum_{{\bf k}}S^f_{{\bf k},i}$,
\bea \label{mixent}
S^f_{{\bf k},i} =- \Big\{(\alpha_{{\bf k},i}^{\dag}\alpha_{{\bf k},i} +\beta_{-{\bf k},i}^{\dag}\beta_{-{\bf k},i} ) \ln{\sin^2{\Gamma_k}}+(\alpha_{{\bf k},i}\alpha_{{\bf k},i}^{\dag} +\beta_{-{\bf k},i}\beta_{-{\bf k},i}^{\dag}  ) \ln{\cos^2{\Gamma_k}} \Big\} ,
\eea
with  $ \sin{\Gamma_k} \equiv |V_{\bf k}|\sin \te $.
We have the following relations - cf. Eq(\ref{relentrTFD}):
\bea
e^{-\frac{S^f_{i}}{2}} \alpha_{{\bf k},j}^{\dag} e^{\frac{S^f_{i} }{2}} = e^{\delta_{ij} \ln\tan{\Gamma_k}}\alpha_{{\bf k},j}^{\dag}\, , \qquad e^{-\frac{S^f_{i}}{2}} \beta_{-{\bf k},j}^{\dag} e^{\frac{S^f_{i} }{2}}=  e^{\delta_{ij} \ln\tan{\Gamma_k}}  \beta_{-{\bf k},j}^{\dag}\,.
\eea
In order to check wether or not the ansatz in Eq.(\ref{entrinv}) is consistent, we evaluate it at the first order approximation in $\theta$ for small $(\Theta_{{\bf k},2}-\Theta_{{\bf k},1})$.
\bea
S^f_{i}\simeq -\sum_{{\bf k}} \Big\{(\alpha_{{\bf k},i}^{\dag}\alpha_{{\bf k},i} +\beta_{-{\bf k},i}^{\dag}\beta_{-{\bf k},i} ) \ln{\theta(\Theta_{{\bf k},2}-\Theta_{{\bf k},1})} \Big\} \,,
\eea
and $ |I_f\rangle \simeq \prod_{{\bf k},r}  \exp\Big\{ \epsilon^{r}\big( \alpha_{{\bf k},1}^{\dag}\beta_{-{\bf k},2}^{\dag} + \alpha_{{\bf k},2}^{\dag}\beta_{-{\bf k},1}^{\dag} ) \Big\}|0\rangle_\mass$, thus the identity in Eq.(\ref{entrinv}) is satisfied in this approximation - cf Eq.(\ref{vacuumaprox1}). This is indeed sufficient for the following considerations. Further discussion on the thermodynamical structure of $|0\rangle_{e, \mu}$ will be presented elsewhere.

Finally, we define the difference $\Delta S^f_{k,i} $ between the vev of the entropy operator Eq.(\ref{mixent}) computed on the two different vacua
\begin{eqnarray}
\Delta S^f_{k,i}= \,_{e, \mu}\langle 0 | S^f_{k,i} | 0\rangle_{e, \mu} - \,_\mass \langle 0 | S^f_{k,i} | 0\rangle_\mass  = -2 \sin^2\Gamma_k\,\ln{\tan^2\Gamma_k }\, . 
\end{eqnarray}
We can now consider the ratio $ \Delta S^f_{k,i}/ \Delta E_{k,i}$ , where the latter is the gap energy defined in the previous Section Eq.(\ref{engap}) for the field $i$, obtaining $ \beta_{k,i}=  \Delta S^f_{k,i} / \Delta E_{k,i} = -  \ln{\tan^2\Gamma_k} / \omega_{ k, i}\,$, which, however, depends on the momentum.
In fact, unlike the standard TFD case, in which the parameter $\gamma_k$ is determined only by the relation in Eq.(\ref{TFDcond}), in the present case the Bogoliubov angle is already set with the condition $\Theta_{{\bf k},i} = \frac {1} {2} \cot^{-1} (\frac{|{\bf k}|}{m_i}$) - see Appendix, Eq.(\ref{condTheta}). This results in an impossibility to introduce a well defined temperature or equivalently in a deviation from the Fermi distribution, due to the non diagonal pairs in the condensate structure of the flavor vacuum.
\begin{figure}[t]
\centering
\includegraphics*[width=12cm]{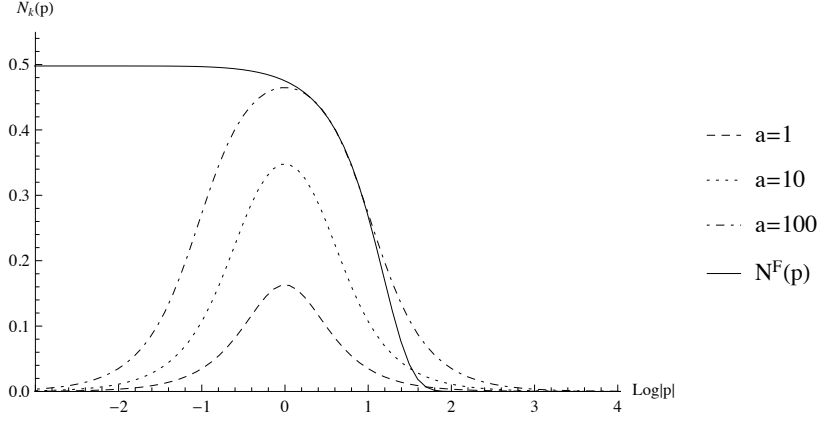}
\caption{Plot of $N_f(p)$ and $N_F(p)$ against $Log |p|$.
 For all curves, we set  $\te = \pi/4$ and $ m_1 = 20$. $N_f(p)$ is plotted for different values of $a$. Sample values of parameters for $N_F(p)$ are $a=100$, $T_1=10^4 $ and $T_2=7.8\,10^4 $.}
\label{FigN}
\end{figure}

On the other hand, starting from a different viewpoint, one can investigate the relation between the flavor vacuum and a thermal vacuum state of the form $|0(\beta_1,\beta_2) \rangle \equiv|0(\beta_1)\rangle  \otimes |0(\beta_2)\rangle$ with
\bea
 |0(\beta_i)\rangle  \equiv \prod_{{\bf k},r} \Big[ \cos{\gamma_{{\bf k},i} (\beta_i)} +\,\sin{\gamma_{{\bf k},i} (\beta_i)}\, \alpha^{r \dagger}_{{\bf k},i} \beta^{r \dagger}_{-{\bf k},i} \Big] |0\ran_{i}\, ,
 \eea
where $ i=1,2$ and $\gamma_{{\bf k},i} (\beta_i)$ are the parameters of the Bogoliubov transformations depending on the temperature.
We recall \cite{BV95} that it is possible to rewrite $|U_\bk|^2$ in terms of two adimensional parameters:
$ |U_\bk|^2 =\Big(1+ 1/ \sqrt{1+a( p/(p^2 +1))^2}\Big) /2 $,
with $ p\equiv \frac {|\bk |}{\sqrt{m_1 m_2}}\, $, $ a\equiv \frac{(m_2 - m_1)^2}{m_1 m_2}\, $.
We consider the total number operator on the flavor vacuum  $N_f(k)\equiv \,_{e, \mu}\lan 0| N_{{\bf k},1}+N_{{\bf k},2} |0\ran_{e, \mu} = 2 \,\sin^2\theta\, |V_{\bf k}|^2  $  while the vev  on the thermal vacuum gives\footnote{F stands for Fermi. We use the notation $ \langle  0(\beta_1,\beta_2) | * |0(\beta_1,\beta_2) \rangle  \equiv  \langle * \rangle_{\beta_1,\beta_2}$.} $N_F(k)\equiv \,\lan N_{{\bf k},1}+N_{{\bf k},2} \rangle_{\beta_1,\beta_2}  = (e^{\beta_1 \omega_{k,1}}+1)^{-1} + (e^{\beta_2 \omega_{k,2}}+1)^{-1} $ .
One may wonder to what extent, $N_F(k)$ can fit $N_f(k)$ for given values of the parameters $m_1$, $m_2$ and $\theta$, by adjusting the free parameters $\beta_1$ and $\beta_2$. From Fig.\ref{FigN} we see that this is somehow possible only for the right tail of the distribution $N_f(k)$; on the other hand, for low momenta, the behavior of the two distributions is quite different.
 This fact boils down to a structural difference between the two states $ | 0\rangle_{e, \mu}$ and $ |0(\beta_1,\beta_2)\rangle $. These states differ because in the condensate structure of the ``thermal'' state $ |0(\beta_1,\beta_2)\rangle $ are missing  terms of the form $({ \al^{r\dag}_{{\bf k},1}\bt^{r\dag}_{-{\bf k},2}}+ { \al^{r\dag}_{{\bf k},2} \bt^{r\dag}_{-{\bf k},1}}) | 0\rangle_\mass$ (cf. Eq.(\ref{flavorvacuum}) of Appendix) due to the non-diagonal Bogoliubov transformation discussed in Section III.

\section{Conclusions}

We have discussed the algebraic structure of the mixing generator for two Dirac neutrino fields with different masses.
We have shown that such a generator can be decomposed in terms of a rotation depending only on the mixing angle and a Bogoliubov transformation depending only on the neutrino masses. These two transformations do not commute among themselves and this fact produces important effects on the vacuum structure.

It is interesting to observe that the  Bogoliubov transformations are indeed responsible for the mass shift and thus the results of this paper can lead to further insight in the interplay between mixing phenomenon and mass generation in a dynamical perspective as recently discussed in Refs.\cite{Blasone:2013qia}.

Moreover, the condensate structure of the vacuum suggests a thermodynamical interpretation which we investigated, showing peculiarities in the thermal behavior due to the  character of the particle-antiparticle condensate involved in the flavor vacuum. Such an issue will be further investigated in a future work.

Finally, we observe that the algebraic mechanism discussed in the present paper appears to be of quite general nature and thus we expect it to hold, with the due differences, also for the mixing among other kinds of fields. For Majorana fields \cite{Blasone:2003hh}, the mixing generator has essentially the same form as the one for Dirac fields Eq.(\ref{genexp}), with the difference that antiparticle ladder operators are replaced by particle operators and  the flavor vacuum appears to be a condensate of pairs of particles with opposite momenta: thus, in such a case, the results here derived apply essentially in the same way, including those concerning the thermodynamical interpretation of \S \,\ref{sect4}. The case of bosonic fields will be discussed in a separate publication together with the extension of the present work to three flavor mixing.

Partial financial support from MIUR and INFN is gratefully acknowledged.

\section*{Appendix}

The fields $\nu_{1}$ and $\nu_{2}$ are expanded as - cf Eq.(\ref{eqn:two})
\be\non
\nu_{i}(x) =  \sum_{r} \int \frac{d^3 \bk}{(2\pi)^\frac{3}{2}}   \lf[u^{r}_{{\bf k},i}(t) \al^{r}_{{\bf k},i}\:+ v^{r}_{-{\bf k},i}(t) \bt^{r\dag }_{-{\bf k},i} \ri] e^{i {\bf k}\cdot {\bf x}}, \;
\quad i=1,2 \;.
\ee
where $u^{r}_{{\bf k},i}(t)=e^{-i\om_{k,i} t}u^{r}_{{\bf k},i}$ and $v^{r}_{{\bf k},i}(t)=e^{i\om_{k,i} t}v^{r}_{{\bf k},i}$, with $\om_{k,i}=\sqrt{{\bf k}^2+m_i^2}$. The $\al^{r}_{{\bf k},i}$ and the $\bt^{r }_{{\bf k},i}$ ($r=1,2$), are the annihilation operators for the vacuum state $|0\ran_{1,2}\equiv|0\ran_{1}\otimes |0\ran_{2}$: $\al^{r}_{{\bf k},i}|0\ran_{1,2}= \bt^{r }_{{\bf k},i}|0\ran_{1,2}=0$.
The anticommutation relations are the standard ones: %
\bea
\label{eqn:three} \{\nu^{\al}_{i}(x),
\nu^{\bt\dag }_{j}(y)\}_{t=t'} = \de^{3}({\bf x}-{\bf y})
\de_{\al\bt} \de_{ij} \;, \;\;\;\;\; \al,\bt=1,..,4 \;,
\\
\label{eqn:four} \{\al^{r}_{{\bf k},i}, \al^{s\dag }_{{\bf q},j}\} =
\de _{{\bf k q}}\de _{rs}\de_{ij}   ;\qquad \{\bt ^{r}_{{\bf k},i},
\bt ^{s\dag}_{{\bf q},j}\} = \de_{{\bf k q}}
\de_{rs}\de_{ij},\;\;\;\; i,j=1,2\;.
\eea
The orthonormality
relations are
$ u^{r\dag}_{{\bf k},i} u^{s}_{{\bf k},i} =   v^{r\dag}_{{\bf k},i}
v^{s}_{{\bf k},i} = \delta_{rs} $ and    $ u^{r\dag}_{{\bf k},i}
v^{s}_{-{\bf k},i} =  v^{r\dag}_{-{\bf k},i} u^{s}_{{\bf k},i} =
0. $ The completeness relation is $\sum_{r}(u^{r}_{{\bf k},i} u^{r\dag}_{{\bf k},i} +
v^{r}_{-{\bf k},i} v^{r\dag}_{-{\bf k},i}) = \ide \;.$

One may recast Eqs.(\ref{fieldmix1}),(\ref{fieldmix2}) as \cite{BV95}:
\bea
\nu_{\si}^{\al}(x) = G^{-1}_{\te}(t)\; \nu_{i}^{\al}(x)\; G_{\te}(t) ,\qquad (\si,i)=(e,1), (\mu,2)
\eea
where the generator $G_{\te}(t)$\footnote{In order to have a simpler notation we will use $G_{\te}(t)\equiv G(t;\te,m_1,m_2)$}  is given by Eq.(\ref{generatorG}). Thus the flavor fields can be expanded as:
\bea\label{eqn:seven}
&&{}\hspace{-1.5cm}
\nu_\si(x)\,=\, \sum_{r=1,2} \int \frac{d^3
\bk}{(2\pi)^\frac{3}{2}} \lf[ u^{r}_{{\bf k},i}(t) \al^{r}_{{\bf k},\si}(t)
+    v^{r}_{-{\bf k},i}(t) \bt^{r\dag}_{-{\bf k},\si}(t) \ri]
e^{i {\bf k}\cdot{\bf x}}\,, \eea
The flavor annihilation operators are defined as $\al^{r}_{{\bf k},\si}(t) \equiv G^{-1}_{\bf \te}(t)\al^{r}_{{\bf k},i} G_{\bf \te}(t)$ etc.
For ${\bf k}=(0,0,|{\bf k}|)$, we have
\bea
\alpha^{r}_{{\bf k},e}(t) & = & \cos\te \alpha^{r}_{{\bf k},1}(t) +  \sin\te \lf( U_{{\bf k}}^*(t)\,\alpha^{r}_{{\bf
k},2}(t) + \epsilon^{r} \,V_{{\bf k}}(t)\beta^{r\dag}_{{-\bf k},2}(t)\ri)
\eea
and similar ones. We have defined
\bea \label{Ut}\hspace{-8mm}
&&U_{{\bf k}}(t)\equiv u^{r\dag}_{{\bf k},2}(t)u^{r}_{{\bf k},1}(t)= v^{r\dag}_{-{\bf k},1}(t)v^{r}_{-{\bf k},2}(t) =\, |U_{{\bf k}}|\,
e^{i(\om_{k,2}-\om_{k,1})t}\,, \\
\label{Vt} \hspace{-8mm}
 &&V_{{\bf k}}(t)\equiv \ep^{r}\;u^{r\dag}_{{\bf k},1}(t)v^{r}_{-{\bf k},2}(t)= -\ep^{r}\;u^{r\dag}_{{\bf k},2}(t)v^{r}_{-{\bf k},1}(t) = \,|V_{{\bf k}}|\;e^{i(\om_{k,2}+\om_{k,1})t}\,.
 \eea
with $\ep^{r}=(-1)^{r}$, and $|U_{{\bf k}}|^{2}+|V_{{\bf k}}|^{2}=1$ with $ |U_{{\bf k}}|=\frac{|{\bf k}|^{2} +(\om_{k,1}+m_{1})(\om_{k,2}+m_{2})}{2 \sqrt{\om_{k,1}\om_{k,2}(\om_{k,1}+m_{1})(\om_{k,2}+m_{2})}} $.

The expansion of the mixing generator in terms of the mass annihilation and creation operators \cite{BV95} is\footnote{In order to simply the notation we omit in the following the time dependance of the annihilation and creation operators}:
\bea
\non
&&\hspace{-8mm}G(\te)=\exp\Biggl\{ \theta \sum_{r}\int\!\frac{d^3 \bk}{(2\pi)^\frac{3}{2}} \Big[U_{\bf k} \lf(\alpha^{r\dag}_{{\bf k},1} \alpha^{r}_{{\bf k},2}  + \beta^{r}_{-{\bf k},1}\beta^{r\dag}_{-{\bf k},2}-\alpha^{r\dag}_{{\bf k},2} \alpha^{r}_{{\bf k},1}-\beta^{r}_{-{\bf k},2}\beta^{r\dag}_{-{\bf k},1}\ri)\\ \label{genexp}
&&\qquad \qquad+\epsilon^{r}V_{\bf k} \lf(\alpha^{r\dag}_{{\bf k},1}\beta^{r\dag}_{-{\bf k},2}- \beta^{r}_{-{\bf k},1}  \alpha^{r}_{{\bf k},2} +\alpha^{r\dag}_{{\bf k},2}\beta^{r\dag}_{-{\bf k},1}  - \beta^{r}_{-{\bf k},2}  \alpha^{r}_{{\bf k},1} \ri)\Big]\Biggr\}\, ,
\eea
Let us now define $ \tilde{R}\equiv \tilde{R}(\te,\Theta_1,\Theta_2) =B^{-1}(\Theta_1,\Theta_2)
R(\te) B(\Theta_1,\Theta_2)\, $ ,
with $B(\Theta_1,\Theta_2)\equiv B_1(\Theta_1) B_2(\Theta_2)$ and $R(\theta)$, $B_i(\Theta_i)$  defined as
in Eqs.(\ref{rotgen}),(\ref{boggen}). $\tilde{R}$ can be written as
\bea
\non
\tilde{R}&=& \exp{\Biggl\{  }  \theta  \sum_r \int \frac{d^3 \bk}{(2\pi)^\frac{3}{2}}
\Big[ \Bigl(\tilde{\alpha}^{r\dagger}_{{\bf k},1}\tilde{\alpha}^{r}_{{\bf k},2}+
\tilde{\beta}^{r\dagger}_{- {\bf k},1} \tilde{\beta}^{r}_{-{\bf k},2} \Bigr) e^{i\psi_k} \\ \label{eqn:twentyone}
\hspace{-5mm}&&\qquad\qquad\qquad \qquad - \Bigl(\tilde{\alpha}^{r\dagger}_{{\bf k},2}
\tilde{\alpha}^{r}_{{\bf k},1} + \tilde{\beta}^{r\dagger}_{-{\bf k},2}
\tilde{\beta}^{r}_{-{\bf k},1} \Bigr) e^{-i\psi_k} \Big]\Biggr\} .
\eea
By use of the explicit form of the Bogoliubov transformed ladder operators, Eqs.(\ref{tialpha}),(\ref{tibeta}) and
imposing the equality between $\tilde{R}$ and $G(\theta)$, we obtain the following conditions for the six parameters (three angles and three phases):
\bea \label{condTheta}
\bar{\Theta}_{{\bf k},i} = \frac {1} {2} \cot^{-1}\lf(\frac{|{\bf k}|}{m_i}\ri)\,, \quad \bar{\phi}_{k,i} = 2 \omega_{k,i} t \, ,\quad  \bar{\psi}_k =  (\omega_{k,1} - \omega_{k,2}) t \, ,\quad  \bar{\theta}=\theta \, ,
\eea
From such constraints, the following relations are derived:
 \bea   U_{\bf k}(t)  = e^{-i\psi_k} \cos(\Theta_{{\bf k},1} -\Theta_{{\bf k},2}) \,, \qquad V_{\bf k}(t)  =  e^{\frac{i(\phi_{k,1} + \phi_{k,2})}{2}}  \sin(\Theta_{{\bf k},1} - \Theta_{{\bf k},2})\,.
 \eea
 In definitive, we have decomposed the mixing generator in the following way\footnote{We used the notation $f(\Theta_i(m_i))\equiv f(m_i)$. In fact $\Theta_{{\bf k},i}$ are functions of the masses and the momentum only. Thus we can regard the generator $B(\Theta_1,\Theta_2)$, where the momentum has been integrated out, as dependent on the mass parameters, i.e. as $B(m_1,m_2)$. }
\bea \label{gendec2}
G(t;\te,m_1,m_2)= B^{-1}(t;m_1,m_2)\,  R(t;\te) \, B(t;m_1,m_2)\, ,
\eea
i.e., as a product of operators depending only on the masses or on the mixing angle.
It is, indeed, possible to disentangle the two dependances, mass and angle, of the mixing generator.
Moreover, the form of the flavor vacuum (at $t \ne 0$) is the following one for ${\bf k}= (0,0,|{\bf k}|)$:
\begin{eqnarray}\non
\label{flavorvacuum}
&&{}\hspace{-9mm}
|0\ran_\flav= \prod_{{\bf k},r} \Bigl[(1-\sin^2\te |V_{{\bf k}}|^2) - \epsilon^r\sin\te\cos\te |V_{{\bf k}}| e^{\frac{i(\phi_1+\phi_2)}{2}}({ \al^{r\dag}_{{\bf k},1}\bt^{r\dag}_{-{\bf k},2}}+{ \al^{r\dag}_{{\bf k},2} \bt^{r\dag}_{-{\bf k},1}}) \\ [1mm]
{}\hspace{-25mm}&& \non
  + \epsilon^r\sin^2\te |V_{{\bf k}}| |U_{{\bf k}}| ({e^{i \phi_2}  \al^{r\dag}_{{\bf k},1}\bt^{r\dag}_{-{\bf k},1}} - { e^{i \phi_1} \al^{r\dag}_{{\bf k},2}\bt^{r\dag}_{-{\bf k},2}} ) \\[1mm] {}\hspace{-25mm}
&{}&+ \sin^2\te  |V_{{\bf k}}|^2 e^{i(\phi_1+\phi_2)}{ \al^{r\dag}_{{\bf k},1}\bt^{r\dag}_{-{\bf k},2} \al^{r\dag}_{{\bf k},2}\bt^{r\dag}_{-{\bf k},1}}\Bigr]|0\ran_\mass\, .
\end{eqnarray}

\smallskip

We report some useful  relations among spinors  of different masses:
\bea
u^r_{{\bf k},1}(t) &=&   u^r_{{\bf k},2}(t) U_{{\bf k}}(t) + \epsilon^r v^r_{-{\bf k},2}(t)  V^*_{{\bf k}}(t)  \\[2mm]
v^r_{-{\bf k},1}(t)  &=&  v^r_{-{\bf k},2}(t) U^*_{{\bf k}}(t)  - \epsilon^r u^r_{{\bf k},2}(t)  V_{{\bf k}}(t)  \\[2mm]
u^r_{{\bf k},2}(t) &=&   u^r_{{\bf k},1}(t)  U^*_{{\bf k}}(t) + \epsilon^r v^r_{-{\bf k},1}(t)  V^*_{{\bf k}}(t)  \\[2mm]
v^r_{-{\bf k},2}(t)  &=&  v^r_{-{\bf k},1}(t) U_{{\bf k}}(t)  - \epsilon^r u^r_{{\bf k},1}(t)  V_{{\bf k}}(t)
\eea
with $U_{{\bf k}}(t)$ and $V_{{\bf k}}(t)$ defined as in Eqs.(\ref{Ut}), (\ref{Vt}). These relations can be easily verified. Consider for example the first one: multiplying on the left by $u^{r\dag}_{{\bf k},2}(t)$, and using the orthonormality relations, we obtain the identity $u^{r\dag}_{{\bf k},2}(t)u^r_{{\bf k},1}(t)=U_{{\bf k}}(t)  $. A similar result is obtained by acting with $v^{r\dag}_{-{\bf k},2}(t)$.

\smallskip

%


\end{document}